\documentclass[a4paper]{llncs}

\usepackage{hyperref}

\usepackage{amsmath}
\usepackage{graphicx}
\usepackage{xspace}
\usepackage{arydshln}
\usepackage{amssymb}
\usepackage{lipsum,longtable}

\usepackage{triotex}
\usepackage[table]{xcolor}

\newcounter{tablerownumbers}
\newcommand\rownumber{\refstepcounter{tablerownumbers}\arabic{tablerownumbers}}

\usepackage{changepage}

\usepackage[font=small,skip=-1pt]{caption}

\usepackage{tikz}
\usetikzlibrary{shapes,arrows,positioning,fit,backgrounds}

\usepackage{times}

\usepackage{multirow}

\usepackage[T1]{fontenc}
\usepackage[scaled=0.81]{beramono} 
\usepackage{eiffel}
\newcommand{\e}[1]{\mbox{\lstinline|#1|}}

\newcommand{\AutoProof}{Au\-to\-Proof\xspace}

\usepackage{boogie}

\newif\ifdraft\drafttrue

\ifdraft
	\newcommand\comm[1]{
		\marginpar{\raggedright\hbadness=10000\scriptsize\it #1\par}}
\else
	\newcommand\comm[1]	{}
\fi

\newif\iflong
\longtrue

\begin{document}

\title{AutoProof: Auto-active Functional \texorpdfstring{\\}{} Verification of Object-oriented Programs}

\author{Julian Tschannen \and Carlo A.\ Furia \and Martin Nordio \and Nadia Polikarpova}

\institute{Chair of Software Engineering, Department of Computer Science, ETH Zurich, Switzerland\\
\email{firstname.lastname@inf.ethz.ch}
}

\maketitle

\begin{abstract}
Auto-active verifiers provide a level of automation intermediate between fully automatic and interactive: users supply code with annotations as input while benefiting from a high level of automation in the back-end.
This paper presents \AutoProof, a state-of-the-art auto-active verifier for object-oriented sequential programs with complex functional specifications.
\AutoProof fully supports advanced object-oriented features and a powerful methodology for framing and class invariants, which make it applicable in practice to idiomatic object-oriented patterns.
The paper focuses on describing \AutoProof's interface, design, and implementation features, and demonstrates \AutoProof's performance on a rich collection of benchmark problems. 
The results attest \AutoProof's competitiveness among tools in its league on cutting-edge functional verification of object-oriented programs.
\end{abstract}

\section{Auto-active Functional Verification of Object-oriented Programs} \label{sec:intro}

Program verification techniques differ wildly in their degree of automation and, correspondingly, in the kinds of properties they target.
One class of approaches---which includes techniques such as abstract interpretation and model checking---is fully \emph{automatic} or ``push button'', the only required input being a program to be verified; to achieve complete automation, these approaches tend to be limited to verifying simple or implicit properties such as absence of invalid pointer dereference.
At the other end of the spectrum are \emph{interactive} approaches to verification---which include tools such as KeY~\cite{KeY}---where the user is ultimately responsible for providing input to the prover on demand, whenever it needs guidance through a successful correctness proof; in principle, this makes it possible to verify arbitrarily complex properties, but it is approachable only by highly-trained verification experts.

In more recent years a new class of approaches have emerged that try to achieve an intermediate degree of automation in the continuum that goes from automatic to interactive---hence their designation~\cite{LeinoMoskal:UsableProgramVerification} as the portmanteau \emph{auto-active}\footnote{Although \emph{inter-matic} would be as good a name.}.
Auto-active tools need no user input during verification, which proceeds autonomously until it succeeds or fails; however, the user is still expected to provide guidance indirectly through \emph{annotations} (such as loop invariants) in the input program.
The auto-active approach has the potential to better support \emph{incrementality}: proving simple properties would require little annotations and of the simple kinds that novice users may be able to provide; proving complex properties would still be possible 
by sustaining a heavy annotation burden.

This paper describes \AutoProof, an auto-active verifier for functional properties of (sequential) object-oriented programs.
In its latest development state, \AutoProof offers a unique combination of features that make it a powerful tool in its category and a significant contribution to the state of the art.
\AutoProof targets a real complex object-oriented programming language (Eiffel)---as opposed to more abstract languages designed specifically for verification.
It supports most language constructs
, as well as a full-fledged verification methodology for heap-manipulating programs based on a flexible annotation protocol, sufficient to completely verify a variety of programs 
that are representative of object-oriented idioms as used in practice. 
\AutoProof was developed with extensibility in mind: 
its annotation library can be augmented with new abstract models, and 
its implementation can 
accommodate changes in the input language.
While Eiffel has a much smaller user base than other object-oriented languages such as 
C++, Java, and C\#, the principles behind \AutoProof are largely language independent; hence, they are relevant to a potentially large number of researchers and users---for whom this paper is written.

The verification challenges we use to evaluate \AutoProof (\autoref{sec:evaluation}) are emerging as the gold standard \cite{vc-vstte10} to demonstrate the capabilities of program provers for functional correctness which, unlike fully automatic tools, use different formats and conventions for input annotations and support specifications of disparate expressiveness, and hence cannot directly be compared on standard benchmark implementations.

Previous work of ours, summarized in \autoref{sec:our-prev-work}, described the individual techniques available in \AutoProof. 
This paper focuses on presenting \AutoProof's functionalities (\autoref{sec:usage}), on describing significant aspects of its design and implementation (\autoref{sec:impl}), and on outlining the results of experiments with realistic case studies, with the goal of showing that \AutoProof's features and performance demonstrate its competitiveness among other tools in its league---auto-active verifiers for object-oriented programs.

\AutoProof is available as part of the open-source Eiffel Verification Environment (EVE) as well as online in your browser; the page 
\begin{center}
\url{http://se.inf.ethz.ch/research/autoproof/} 
\end{center}
contains source and binary distributions, detailed usage instructions, a user manual, an interactive tutorial, and the benchmarks solutions discussed in \autoref{sec:evaluation}.

\section{Related Work} \label{sec:related}

\subsection{Program Verifiers}
In reviewing related work, we focus on the tools that are closer to \AutoProof in terms of features, and design principles and goals.
Only few of them are, like \AutoProof, auto-active, work on real object-oriented programming languages, and support the verification of general functional properties.
Krakatoa~\cite{Krakatoa} belongs to this category, as it works on Java programs annotated with a variant of JML (the Java Modeling Language~\cite{JML}).
Since it lacks a full-fledged methodology for class invariants and framing, using Krakatoa to verify object-oriented idiomatic patterns---such as those we discuss in \autoref{sec:problems}---would be quite impractical; in fact, the reference examples distributed with Krakatoa target the verification of algorithmic problems where object-oriented features are immaterial.
Similar observations apply to the few other auto-active tools working on Java and JML, such as ESC/Java2~\cite{ESCJava2} 
or the more recent OpenJML~\cite{OpenJML,OpenJMLPaper}.
Even when ESC/Java2 was used on industrial-strength case studies (such as the KOA e-voting system~\cite{KOA}), the emphasis was on modeling and correct-by-construction development, and verification was normally applied only to limited parts of the systems.
By contrast, the Spec\# system~\cite{SpecSharp} was the forerunner in a new research direction, also followed by \AutoProof, that focuses on the complex problems raised by object-oriented structures with sharing, object hierarchies, and collaborative patterns.
Spec\# works on an annotation-based dialect of the C\# language and supports an ownership model which is suitable for hierarchical object structures, as well as visibility-based invariants to specify more complex object relations.
Collaborative object structures as implemented in practice (\autoref{sec:problems}) require, however, 
more flexible methodologies~\cite{semicola} not currently available in Spec\#.
Tools, such as VeriFast~\cite{JacobsSmansPiessens}, based on separation logic provide powerful methodologies through abstractions different than class invariants, which may lead to a lower level of automation than tools such as \AutoProof and a generally higher annotation overhead---ultimately targeting highly trained users.

The experience with the Spec\# project suggested  that targeting a real object-oriented programming language introduces numerous complications and may divert the focus away from fundamental problems in tool-supported verification.
The Dafny program verifier~\cite{Dafny} was developed based on this lesson: it supports a simple language expressly designed for verification, which eschews most complications of real object-oriented programming languages (such as inheritance and a complex memory model). 
Other auto-active verifiers target programming language paradigms other than object orientation.
Leon~\cite{Leon} and Why3~\cite{Why3}, for example, work on functional programming languages---respectively, a subset of Scala and a dialect of ML; VCC~\cite{VCC} works on C programs and supports object invariants but with an emphasis on memory safety of low-level concurrent code.

\AutoProof lies between automatic and interactive tools in the wide spectrum of verification tools.
The CodeContract checker (formerly known as Clousot~\cite{Clousot}) is a powerful static analyzer for .NET languages that belongs to the former category (and hence it is limited to properties expressible in its abstract domains).
The KeY system~\cite{KeY} for Java belongs to the latter category: while it supports SMT solvers as back-ends to automatically discharge some verification conditions, its full-fledged usage requires explicit user interactions to guide the prover through the verification process.

\subsection{Our Previous Work on \AutoProof} 
\label{sec:our-prev-work}
In previous work, we formalized some critical object-oriented features as they are available in Eiffel, notably function objects 
(called ``agents'' in Eiffel) and inheritance and polymorphism~\cite{laser-paper}.
An important aspects for usability is reducing annotation overhead; to this end, we introduced heuristics known as ``two-step verification''~\cite{two-step-vstte} and demonstrated them on algorithmic challenges~\cite{sttt-paper}.
We recently presented the theory behind \AutoProof's invariant methodology~\cite{semicola}, which includes full support for class invariants, framing, and ghost code. 
The current paper discusses how these features are available in \AutoProof, with a focus on advanced object-oriented verification challenges.

\section{Using AutoProof} \label{sec:usage}

\AutoProof is a static verifier for Eiffel programs which interacts with users according to the auto-active paradigm~\cite{LeinoMoskal:UsableProgramVerification}: verification attempts are completely automated (``push button''), but users are expected in general to provide additional information in the form of \emph{annotations} (loop invariants, intermediate assertions, etc.) for verification to succeed.

\AutoProof targets the verification of functional correctness.
Given a collection of Eiffel classes, it tries to establish that: routines satisfy their pre/post and frame specifications and maintain class invariants; routine calls take place in states satisfying the callee's precondition; loops and recursive calls terminate; integer variables do not overflow; there are no dereferences of \e{Void} (\B{null}) objects.

\AutoProof's techniques are \emph{sound}\footnote{As usual, modulo bugs in the implementation.}: successful verification entails that the input program is correct with respect to its given specification.
Since it deals with expressive specifications, \AutoProof is necessarily \emph{incomplete}: failed verification may indicate functional errors but also shortcomings of the heuristics of the underlying theorem prover (which uses such heuristics to reason in practice about highly-complex and undecidable logic fragments).

Dealing with inconclusive error reports in incomplete tools is a practical hurdle to usability that can spoil user experience---especially for novices.
To improve user feedback in case of failed verification attempts, \AutoProof implements a collection of heuristics known as ``\emph{two-step verification}''~\cite{two-step-vstte}.
When they are enabled, each failed verification attempt is transparently followed by a second step that is in general unsound (as it uses under-approximations such as loop unrolling) but helps discern whether failed verification is due to real errors or just to insufficiently detailed annotations.
Users see the combined output from the two steps in the form of suggestions to improve the program and its annotations.
For example, if verification of a loop fails in the first step but succeeds with finite loop unrolling, 
the suggestion is that there are no obvious errors in the loop but the loop invariant should be strengthened to make it inductive.

\subsection{User Interface (UI)} \label{sec:ui}

\begin{figure}[!tb]
\begin{center}
\includegraphics[width=0.85\columnwidth]{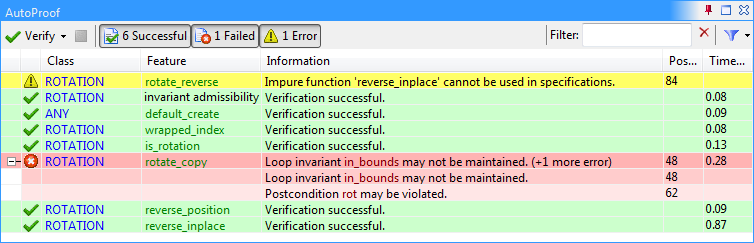}
\end{center}
\caption{The \AutoProof output panel showing verification results in EVE.}
\label{fig:eve-tool-result}
\end{figure}

\AutoProof offers its core functionalities both through a command line interface (CLI) and a library (API).
End users normally interact with \AutoProof through one of two graphical interfaces (GUI): a web-based GUI is available at \url{http://cloudstudio.ethz.ch/comcom/#AutoProof}; and \AutoProof is fully integrated in EVE, the open-source research branch of the EiffelStudio development environment.
The following presentation focuses on \AutoProof in EVE, but most features are available in every UI.

Users launch \AutoProof on the current project, or on specific classes or members thereof.
Verification proceeds in the background until it terminates, is stopped, or times out. 
 Results are displayed in a panel such as in \autoref{fig:eve-tool-result}: each entry corresponds to a routine of some class and is colored to summarize verification outcome.
Green entries are successfully verified; red entries have failed verification; and yellow entries denote invalid input, which cannot be translated and verified (for example, 
impure functions with side effects used as specification elements determine invalid input).
Red entries can be expanded into more detailed error messages or suggestions to fix them (when enabled, two-step verification helps provide more precise suggestions).
For example, the failed verification entry for a routine may detail that its loop invariant may not be maintained, or that it may not terminate; and suggest that the loop invariant be strengthened, or a suitable variant be provided.

\AutoProof's UI is deliberately kept simple with few options and sensible defaults.
For advanced users, fine-grained control over \AutoProof's behavior is still possible through program annotations, which we outline in the next section.

\subsection{Input Language Support} \label{sec:input-language}

\AutoProof supports most of the Eiffel language as used in practice, obviously including Eiffel's native notation for contracts (specification elements) such as pre- and postconditions, class invariants, loop invariants and variants, and inlined assertions 
such as \e{check} (\B{assert} in other languages). 
Object-oriented features---classes and types, multiple inheritance, polymorphism---are fully sup\-port\-ed~\cite{laser-paper}, and so are imperative and procedural constructs.

\textbf{Partially supported and unsupported features.}
A few language features that \AutoProof does not currently fully support have a semantics that violates well-formedness conditions required for verification: 
AutoProof doesn't support specification expressions with side effects (for example, a precondition that creates an object). 
It also doesn't support the semantics of \e{once} routines (similar to \lstinline[language=Java]|static| in Java and C\#), which would require global reasoning thus breaking modularity.

Other partially supported features originate in the distinction between machine and mathematical representation of types.
Among primitive types, machine \e{INTEGER}s are fully supported (including overflows); floating-point \e{REAL}s are modeled as infinite-precision mathematical reals; strings are not supported but for single-character operations.
Array and list library containers with simplified interfaces are supported out of the box.
Other container types require custom specification; we recently developed a fully verified full-fledged data structure library including sets, hash tables, and trees~\cite{EB2}. 
Agents (function objects) are partially supported, with some restrictions in their specifications~\cite{laser-paper}.
The semantics of native \e{external} routines 
is reduced to their specification.
We designed~\cite{laser-paper} a translation for exceptions based on the latest draft of the Eiffel language standard, but \AutoProof doesn't support it yet since the Eiffel compiler still only implements the obsolete syntax for exceptions (and exceptions have very limited usage in Eiffel anyway).

\textbf{Annotations for verification.}
Supporting effective auto-active verification requires much more than translating the input language and specification into verification conditions.
\AutoProof supports \emph{semantic collaboration}, a full-fledged framing methodology we designed to reason about class invariants of structures made of collaborating objects, integrated with a standard \emph{ownership} model; both are described in detail in our previous work~\cite{semicola}.
\AutoProof's verification methodology relies on annotations that are not part of the Eiffel language.
Annotations in assertions or other specification elements use predefined dummy features with empty implementation. 
Annotations of this kind include \emph{modify} and \emph{read} clauses (specifying objects whose state may be modified or read by a routine's body). 
For instance, a clause \e{modify (set)} in a routine's precondition denotes that executing the routine may modify objects in \e{set}.

Annotations that apply to whole classes or features are expressed by means of Eiffel's \e{note} clauses, which attach additional information that is ignored by the Eiffel compiler but is processed by \AutoProof.
Annotations of this kind include defining class members as \emph{ghost} (only used in specifications), procedures as \emph{lemmas} (outlining a proof using assertions and ghost-state manipulation), and which members of a class define its abstract \emph{model} (to be referred to in interface specifications).
For example \e{note status: ghost} tags as ghost the member it is attached to.

A distinctive trait of semantic collaboration, as available to \AutoProof users, is the combination of flexible expressive annotations with useful defaults.
Flexible annotations offer fine-grained control over the visibility of specification elements (for example, invariant clauses can be referenced individually); defaults reduce the amount of required manual annotations in many practical cases.
The combination of the two is instrumental in making \AutoProof usable on complex examples of realistic object-oriented programs.

\textbf{Verifier's options.}
\AutoProof \emph{verification options} are also expressed by means of \e{note} clauses: users can disable generating boilerplate implicit contracts, skip verification of a specific class, 
disable termination checking (only verify partial correctness), and define a custom mapping of a class's type to a Boogie theory file.
See \AutoProof's manual for a complete list of features, options, and examples of usage.

\textbf{Specification library.}
To support writing complex specifications, \AutoProof provides a library---called MML for Mathematical Model Library---of pre-defined abstract types .
These includes mathematical structures such as sets, relations, sequences, bags (multisets), and maps.
The MML annotation style follows the model-based par\-a\-digm~\cite{PFM10-VSTTE10}, which helps write abstract and concise, yet expressive, specifications.
MML's features are fully integrated in \AutoProof by means of effective mappings to Boogie background theories.
A distinctive advantage of providing mathematical types as an annotated library is that MML is \emph{extensible}: users can easily provide additional abstractions by writing annotated Eiffel classes and by linking them to background theories using custom \e{note} annotations---in the very same way existing MML classes are defined.
This is not possible in most other auto-active verifiers, where mathematical types for specification are built into the language syntax.

\begin{figure}[!hbt]
\begin{lstlisting}[basicstyle=\scriptsize,xleftmargin=15mm]
	binary_search (a: ARRAY [INTEGER]; value: INTEGER): INTEGER
		require sorted: is_sorted (a.sequence)
		local low, up, middle: INTEGER
		do
			from low := 1; up := a.count + 1
			invariant
				low_and_up_range: 1 <= low and low <= up and up <= a.count + 1
				result_range: Result = 0 or 1 <= Result and Result <= a.count
				not_left: across 1 |..| (low-1) as i all a.sequence[i] < value end
				not_right: across up |..| a.count as i all value < a.sequence[i] end
				found: Result > 0 implies a.sequence[Result] = value
			until low >= up or Result > 0
			loop
				middle := low + ((up - low) // 2)
				if a[middle] < value then low := middle + 1
				elseif a[middle] > value then up := middle
				else Result := middle end
			variant (a.count - Result) + (up - low) end
		ensure
			present: a.sequence.has (value) = (Result > 0)
			not_present: not a.sequence.has (value) = (Result = 0)
			found_if_present: Result > 0 implies a.sequence[Result] = value
		end
\end{lstlisting}
\caption{Binary search implementation verified by \AutoProof.}
\label{fig:binsearch}
\end{figure}

\textbf{Input language syntax.}
\autoref{fig:binsearch} shows an example of annotated input: an implementation of binary search (problem \nref{p:binsearch} in \autoref{tab:Problems}) that \AutoProof can verify.
From top to bottom, the routine \e{binary_search} includes signature, precondition (\e{require}), \e{local} variable declarations, body consisting of an initialization (\e{from}) followed by a \e{loop} that executes \e{until} its exit condition becomes true, and postcondition (\e{ensure}).
The loop's annotations include loop \e{invariant} and \e{variant}.
Each specification element consists of clauses, one per line, with a \emph{tag} (such as \e{sorted} for the lone precondition clause) for identification in error reports.
Quantified expressions in contracts use the \e{across} syntax, which corresponds to (bounded) first-order universal (\e{across ... all}) and existential (\e{across ... some}) quantification. 
For example, loop invariant clause \e{not_left} corresponds to $\forall i \colon 1 \leq i < \e{low} \Longrightarrow \e{a.sequence[}i\e{]} < \e{value}$.

\section{How AutoProof Works: Architecture and Implementation} \label{sec:impl}

As it is customary in deductive verification, \AutoProof translates input programs into verification conditions (VCs): logic formulas whose validity entails correctness of the input programs.
Following the approach pioneered by Spec\#~\cite{SpecSharp} and since adopted by numerous other tools, \AutoProof does not generate VCs directly but translates Eiffel programs into Boogie programs~\cite{BoogieManual} and calls the Boogie tool to generate VCs from the latter.
Boogie is a simple procedural language tailored for verification, as well as a verification tool that takes programs written in the Boogie language, generates VCs for them, feeds the VCs to an SMT solver (Z3 by default), and interprets the solver's output in terms of elements of the input Boogie program.
Using Boogie decouples VC generation from processing the source language (Eiffel, in \AutoProof's case) and takes advantage of Boogie's efficient VC generation capabilities.

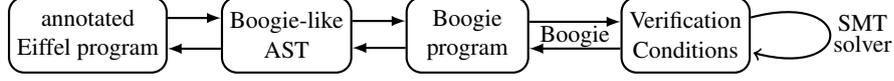
\begin{figure}[!tbh]
\begin{center}
  \begin{tikzpicture}[
  comp/.style={rectangle, minimum width=16mm, minimum height=10mm, thick,rounded corners=2mm,draw=black,font=\footnotesize,align=center},
  node distance=7mm,
  pin distance=7mm,
  every pin edge/.style={<-, shorten <=4mm, ultra thick,green!40!black!60},
  ]
  \lstset{basicstyle=\footnotesize}

  \node (eiffel) [comp] {annotated \\ Eiffel program};
  \node (ir) [comp, right=of eiffel] {Boogie-like \\AST};
  \node (boogie) [comp, right=of ir] {Boogie \\ program};
  \node (vc) [comp, right=12mm of boogie] {Verification \\ Conditions};

  \setlength{\baselineskip}{8pt}

  \begin{scope}[-latex,thick,align=center]
    \path (node cs:name=eiffel,angle=12) edge node {} ({node cs:name=eiffel,angle=12}-|ir.west);
    \path (node cs:name=ir,angle=12) edge node {} ({node cs:name=ir,angle=12}-|boogie.west);
    \path (node cs:name=boogie,angle=12) edge node [below=-2pt] {Boogie} ({node cs:name=boogie,angle=12}-|vc.west);
    \path (node cs:name=vc,angle=-168) edge node {} ({node cs:name=vc,angle=-168}-|boogie.east);
    \path (node cs:name=boogie,angle=-168) edge node {} ({node cs:name=boogie,angle=-168}-|ir.east);
    \path (node cs:name=ir,angle=-168) edge node {} ({node cs:name=ir,angle=-168}-|eiffel.east);
    \path (vc) edge [-latex, loop right] node [above right=-3.5mm and -2pt] {SMT\\solver} (vc);
  \end{scope}
\end{tikzpicture}
\end{center}
\caption{Workflow of \AutoProof with Boogie back-end.}
\label{fig:ap-workflow}
\end{figure}

As outlined in \autoref{fig:ap-workflow}, \AutoProof implements the translation from Eiffel to Boogie in two stages.
In the first stage, it processes an input Eiffel program and translates it into a Boogie-like abstract syntax tree (AST); in the second stage, \AutoProof transcribes the AST into a textual Boogie program.

The rest of this section \iflong outlines the overall mapping from Eiffel to Boogie (\autoref{sec:translation}), and then \fi focuses on describing how \AutoProof's architecture (\autoref{sec:architecture}) and implementation features (\autoref{sec:implementation-features}) make for a flexible and customizable translation process.
\iflong\else An extended version of this paper~\cite{AutoProof-tool-TR-20150114} also outlines the mapping from Eiffel to Boogie. \fi
We focus on discussing the challenges tackled when developing \AutoProof and the advantages of our implemented solutions.

\iflong
\subsection{Eiffel to Boogie Encoding} \label{sec:translation}

We briefly recall the core features of how Eiffel is encoded in Boogie; see our previous work~\cite{laser-paper} for details.
Boogie is a simple procedural language without explicit support for object-oriented features or any form of dynamic memory.
Therefore, a key feature of the encoding is a \emph{memory model} mapping Eiffel's heap, references, and objects.
The \B{Heap} is a Boogie global variable
\begin{boogieNoLines}
var Heap: <<\alpha>> $\;$[ref, Field \alpha] \alpha
\end{boogieNoLines}
which maps pairs of reference (type \B{ref}) and field name (type \B{Field}) into values.

The \emph{heap} type is polymorphic with respect to the value type $\alpha$, which can be any of the available types---including references \B{ref} and Boogie primitive types such as \B{int} and \B{bool}.
In the encoding, Eiffel \emph{reference types} become \B{ref} in Boogie, whereas Eiffel \emph{primitive types} become Boogie primitive types with axiomatic constraints to ensure consistent representation.
For example, Eiffel variables of type \e{INTEGER} map to Boogie variables of type \B{int}; the latter, however, corresponds to mathematical integers whereas the former has 32-bit precision.
To bridge the semantic gap, functions such as
\begin{boogieNoLines}
function is_integer(i: int): (bool) { $- 2^{31} \leq$ i $\leq 2^{31} - 1$ }
\end{boogieNoLines}
are used in the Boogie encoding to check for absence of overflows in operations that manipulate Eiffel integers.

An Eiffel \emph{class declaration} \e{class C inherit B feature a: A end} introduces a class \e{C}, inheriting from another class \e{B}, that includes an attribute \e{a} of type \e{A}. 
The Boogie encoding of \e{C} comprises an uninterpreted type identifier \mbox{\B{const unique C: Type}} related, in an axiom, to \e{B}'s type identifier by the ``subtype of'' partial order relation \mbox{$\MB{C} <\colon \MB{B}$.}
The axioms about classes are built around this relation so as to support polymorphism.

\e{C}'s \emph{attribute} \e{a} determines a field name \B{const C.a: Field ref}, which is used to access the object attached to \e{a} in the heap.
For example, the Eiffel assignment \e{c.a := x} (where \e{c} is a reference of class \e{C} and \e{a} is one of \e{C}'s attributes) is expressed in Boogie as \B{Heap := Heap[c, C.a := x]}: update to \B{x}  the mapping of the pair $(\MB{c}, \MB{C.a})$ in global variable \B{Heap}, while leaving the rest of the mapping unchanged.

Eiffel routines translate to Boogie \B{procedure}s: Eiffel procedures (not returning any value) update the heap according to their modify clauses; Eiffel functions are assumed pure by default, in which case an Eiffel function \e{f} of class \e{C} translates to both a Boogie \B{procedure C.f} and a Boogie \B{function fun.C.f}. 
The former's postcondition includes a check that \e{C.f} returns values consistent with \B{fun.C.f}'s definition.
Then, \e{C.f} is used to check \e{f}'s correctness against its specification and to reason about usages of \e{f} in imperative code; \e{fun.C.f} encodes \e{f}'s values when they're used in specification elements.

The translation of \emph{control-flow} structures (conditionals and loops) uses the corresponding Boogie structures whenever possible.
The encoding of advanced verification features such as semantic collaboration and two-step verification is described in related work~\cite{laser-paper,semicola,two-step-vstte} (see also \autoref{sec:input-language}).
\fi

\subsection{Extensible Architecture} \label{sec:architecture}

\textbf{Top-level API.}
Class \e{AUTOPROOF} is the main entry point of \AutoProof's API.
It offers features to submit Eiffel code, and to start and stop the verification process.
Objects of class \e{RESULT} store the outcome of a verification session, which can be queried by calling routines of the class.
One can also register an Eiffel \e{agent} (function object) with an \e{AUTOPROOF} object; the outcome \e{RESULT} object is passed to the agent for processing as soon as it is available.
This pattern is customary in reactive applications such as \AutoProof's GUI in EVE.

\textbf{Translation to Boogie.}
An abstract syntax tree (AST) expresses the same semantics as Eiffel source code but using elements reflecting Boogie's constructs.
Type relations such as inheritance are explicitly represented (based on type checking) using axiomatic constraints, so that ASTs contain all the information necessary for verification.
The transcription of the AST into a concrete Boogie program is implemented by a \emph{visitor}\iflong~\cite{Gof:1995}\fi{} of the AST. 
Modifying \AutoProof in response to changes in Boogie's syntax would only require to modify the visitor. 

\textbf{Extension points.}
\AutoProof's architecture incorporates \emph{extension points} where it is possible to programmatically modify and extend \AutoProof's behavior to implement different verification processes.
Each extension point maintains a number of \emph{handlers} that take care of aspects of the translation from Eiffel to the Boogie-like AST.
Multiple handlers are composed according to the \emph{chain of responsibility} pattern\iflong~\cite{Gof:1995}\fi; this means that a handler may only implement the translation of one specific source language element, while delegating to the default \AutoProof handlers in all other cases.
A new translation feature can thus be added by writing a handler and registering it at an extension point.
Extension points target three program elements of different generality.
\begin{description}
\item[Across] extension points handle the translation of Eiffel \e{across} expressions, which correspond to quantified expressions. Handlers can define a semantics of quantification over arbitrary data structures and domains. (\AutoProof uses this extension point to translate quantifications over arrays and lists.)
\item[Call] extension points handle the translation of Eiffel calls, both in executable code and specifications. Handlers can define translations specific to certain data types. (\AutoProof uses this extension point to translate functions on integers and dummy features for specification.)
\item[Expression] extension points handle the translation of expressions. Handlers can define translations of practically every Eiffel expression into a Boogie-like AST representation. This extension point subsumes the other two, which offer a simpler interface sufficient when only specific language 
elements require a different translation. 
\end{description}
The flexibility provided for by extension points is particular to \AutoProof: the architecture of other similar tools (Spec\#, Dafny, and OpenJML) does not seem to offer comparable architectural features for straightforward extensibility in the object-oriented style.

\subsection{Implementation Features} \label{sec:implementation-features}

\AutoProof's implementation consists of about 25'000 lines of Eiffel code in 160 classes.

\textbf{Modular translation.}
\AutoProof performs \emph{modular} reasoning: the effects of a call to \e{p} within routine \e{r}'s body are limited to what is declared in \e{p}'s specification (its pre- and postcondition and frame) irrespective of \e{p}'s body (which is only used to verify \e{p}'s correctness).
To achieve modularity incrementally, \AutoProof maintains a \emph{translation pool} of references to Eiffel elements (essentially, routines and their specifications).
Initially, it populates the pool with references to the routines of the classes specified as input to be verified.
Then, it proceeds as follows: (1) select an element \e{el} from the pool that hasn't been translated yet; (2) translate \e{el} into Boogie-like AST and mark \e{el} as translated; (3) if \e{el} refers to (i.e., calls) any element \e{p} not in the pool, add a reference to \e{p}'s specification to the pool; (4) if all elements in the pool are marked as translated stop, otherwise repeat (1).
This process populates the pool with the transitive closure of the ``calls'' relation, whose second elements in relationship pairs are specifications, starting from the input elements to be verified.

\textbf{Traceability of results.}
The auto-active paradigm is based on interacting with users at the high level of the source language; in case of failed verification, 
reports must 
refer to 
the input Eiffel program rather than to the lower level (Boogie code). To this end, \AutoProof follows the standard approach of adding structured comments to various parts of the Boogie code---most importantly to every assertion that undergoes verification: postconditions; preconditions of called routine at call sites; loop invariants; and other intermediate \B{assert}s. Comments may include information about the \emph{type} of condition that is checked (postcondition, loop termination, etc.), the \emph{tag} identifying the clause (in Eiffel, users can name each assertion clause for identification), a \emph{line} number in the Eiffel program, the \emph{called} routine's name (at call sites), and whether an assertion was \emph{generated} by applying a default schema that users have the option to disable (such as in the case of default class invariant annotations~\cite{semicola}).
For each assertion that fails verification, \AutoProof reads the information in the corresponding comment and makes it available in a \e{RESULT} object to the \e{agent}s registered through the API to receive verification outcomes about some or all input elements.
\e{RESULT} objects also include information about verification times.
This \emph{publish/subscribe} scheme provides fine-grained control on how results are displayed.

\textbf{Bulk vs.\ forked feedback.}
\AutoProof provides feedback to users in one of two modes.
In \emph{bulk} mode all input is translated into a single Boogie file; results are fed back to users when verification of the whole input has completed.
Using \AutoProof in bulk mode minimizes translation and Boogie invocation overhead but provides feedback synchronously, only when the whole batch has been processed.
In contrast, \AutoProof's \emph{forked} mode offers asynchronous feedback: each input routine (and implicit proof obligations such as for class invariant admissibility checking) is translated into its own self-contained Boogie file; parallel instances of Boogie run on each file and results are fed back to users asynchronously as soon as any Boogie process terminates.
\AutoProof's UIs use the simpler bulk mode by default, but offer an option to switch to the forked mode when responsiveness and a fast turnaround are deemed important.

\section{Benchmarks and Evaluation} \label{sec:evaluation}

We give capsule descriptions of benchmark problems that we verified using the latest version of \AutoProof; the complete solutions are available at \url{http://se.inf.ethz.ch/research/autoproof/repo} through \AutoProof's web interface.

\subsection{Benchmarks Description} \label{sec:problems}

Our selection of problems is largely based on the verification challenges put forward during several scientific forums, namely the SAVCBS workshops~\cite{SAVCBS}, and various verification competitions~\cite{vc-vstte10,vc-cost11,vc-vstte12,vc-fm12} and benchmarks~\cite{vsi-benchmarks}.
These challenges have recently emerged as the customary yardstick against which to measure progress and open challenges in verification of full functional correctness.

\autoref{tab:Problems} presents a short description of verified problems. 
For complete descriptions see the references (and \cite{semicola} for our solutions to problems \ref{p:iterator}--\ref{p:pip}).
The table is partitioned in three groups: the first group (\ref{p:arith}--\ref{p:sorting}) includes mainly \emph{algorithmic} problems; the second group (\ref{p:iterator}--\ref{p:pip}) includes object-oriented design challenges that require complex \emph{invariant} and \emph{framing} methodologies; the third group (\ref{p:closures}--\ref{p:game2}) targets \emph{data-structure} related problems that combine algorithmic and invariant-based reasoning.
The second and third group include cutting-edge challenges of reasoning about functional properties of objects in the heap; for example, \nref{p:pip} describes a data structure whose node invariants depend on objects not accessible in the physical heap.

\begin{table}
\begin{center}
\begin{scriptsize}
\setlength{\tabcolsep}{3pt}


\begin{tabular}{rl l r}
\textsc{\#} & \textsc{name} &  \textsc{description} 
 & \textsc{from}  \\
\hline

\rownumber \namelabel{p:arith}{\textsc{arith}} &
Arithmetic (\nref{p:arith}) 
&
Build arithmetic operations based on the increment operation.
&
\cite{vsi-benchmarks}	 
\\

\rownumber \namelabel{p:binsearch}{\textsc{bins}} &
Binary search (\nref{p:binsearch}) 
&
Binary search on a sorted array (iterative and recursive version).
&
\cite{vsi-benchmarks}	
\\

\rownumber \namelabel{p:sum-max}{\textsc{s\&m}} &
Sum \& max (\nref{p:sum-max}) &
Sum and maximum of an integer array.
&
\cite{vc-vstte10}	
\\

\rownumber \namelabel{p:search-list}{\textsc{search}} &
Search a list (\nref{p:search-list})
&
Find the index of the first zero element in a linked list of integers.
&
\cite{vc-vstte10}	
\\

\rownumber \namelabel{p:twaymax}{\textsc{2-max}} &
Two-way max (\nref{p:twaymax}) &
Find the maximum element in an array by searching at both ends.
&
\cite{vc-cost11}	
\\

\rownumber \namelabel{p:twaysort}{\textsc{2-sort}} &
Two-way sort (\nref{p:twaysort}) &
Sort a Boolean array in linear time using swaps at both ends.
&
\cite{vc-vstte12}	
\\

\rownumber \namelabel{p:dutchflag}{\textsc{dutch}} &
Dutch flag (\nref{p:dutchflag}) &
Partition an array in three different regions (specific and general verions).
&
\cite{Dij76}	
\\

\rownumber \namelabel{p:lcp}{\textsc{lcp}} &
LCP (\nref{p:lcp}) &
Longest common prefix starting at given positions $x$ and $y$ in an array.
 &
\cite{vc-fm12}	
\\

\rownumber \namelabel{p:rotation}{\textsc{rot}} &
Rotation (\nref{p:rotation}) &
Circularly shift a list by $k$ positions (multiple algorithms).
&
\cite{rotation}	
\\

\rownumber \namelabel{p:sorting}{\textsc{sort}} &
Sorting (\nref{p:sorting}) &
Sorting of integer arrays (multiple algorithms).
&

\\

\cline{1-2}

\rownumber \namelabel{p:iterator}{\textsc{iter}}  &
Iterator (\nref{p:iterator}) & 
Multiple iterators over a collection are invalidated when the content changes.
& 
\cite['06]{SAVCBS}
\\

\rownumber \namelabel{p:subj-obs}{\textsc{s/o}}  &
Subject/observer (\nref{p:subj-obs}) &
Design pattern: multiple observers cache the content of a subject object.
 &
\cite['07]{SAVCBS}
\\

\rownumber \namelabel{p:composite}{\textsc{cmp}} &
Composite (\nref{p:composite}) &
Design pattern: a tree with consistency between parent and children nodes.
&
\cite['08]{SAVCBS}
\\

\rownumber \namelabel{p:mclock}{\textsc{mc}} &
Master clock (\nref{p:mclock}) &
A number of slave clocks are loosely synchronized to a master.
&
\cite{Friends}
\\

\rownumber \namelabel{p:marriage}{\textsc{mar}} &
Marriage (\nref{p:marriage}) &
Person and spouse objects with co-dependent invariants.
&
\cite{Dynamic}
\\

\rownumber \namelabel{p:dllist}{\textsc{dll}} &
Doubly-linked list (\nref{p:dllist}) &
Linked list whose nodes have links to left and right neighbors.
&
\cite{Dynamic}
\\

\rownumber \namelabel{p:pip}{\textsc{pip}} &
PIP (\nref{p:pip}) &
Graph structure with cycles where each node links to at most one parent.
&
\cite{Need} 
\\

\cline{1-2}

\rownumber \namelabel{p:closures}{\textsc{close}} &
Closures (\nref{p:closures}) &
Various applications of function objects.
&
\cite{facj07}	
\\
 
\rownumber \namelabel{p:strategy}{\textsc{strat}} &
Strategy (\nref{p:strategy}) &
Design pattern: a program's behavior is selected at runtime. &
\cite{facj07}
\\

\rownumber \namelabel{p:cmd}{\textsc{cmd}} &
Command (\nref{p:cmd}) &
Design pattern: encapsulate complete information to execute a command. &
\cite{facj07}
\\

\rownumber \namelabel{p:map}{\textsc{map}} &
Map ADT (\nref{p:map}) 
&	
Generic map ADT with layered data.
&
\cite{vsi-benchmarks}	
\\

\rownumber \namelabel{p:linked-queue}{\textsc{queue}} &
Linked queue (\nref{p:linked-queue}) &	
Queue implemented using a linked list.
&
\cite{vsi-benchmarks}	
\\

\rownumber \namelabel{p:tree-max}{\textsc{tmax}} &
Tree maximum (\nref{p:tree-max}) &
Find the maximum value in nodes of a binary tree.
&
\cite{vc-cost11}	
\\

\rownumber \namelabel{p:ringbuffer}{\textsc{buff}} &
Ring buffer (\nref{p:ringbuffer}) &
A bounded queue implemented using a circular array.
&
\cite{vc-vstte12}
\\

\rownumber \namelabel{p:hashset}{\textsc{hset}} &
Hash set (\nref{p:hashset}) &
A hash set with mutable elements.
&
\\

\rownumber \namelabel{p:game1}{\textsc{game1}} &
Board game 1 (\nref{p:game1}) &
A simple board game application: players throw dice and move on a board.
&
\\

\rownumber \namelabel{p:game2}{\textsc{game2}} &
Board game 2 (\nref{p:game2}) &
A more complex board game application: different board-square types.
&
\\

\end{tabular}
\end{scriptsize}
\end{center}
\caption{Descriptions of benchmark problems.}
\label{tab:Problems}
\end{table}

\subsection{Verified Solutions with AutoProof} \label{sec:solutions}

\autoref{tab:Solutions} displays data about the verified solutions to the problems of \autoref{sec:problems}; for each problem: the number of Eiffel classes (\textsc{\#C}) and routines (\textsc{\#R}), the latter split into \textit{gh}ost functions and lemma procedures and \textit{co}ncrete (non-ghost) routines; the lines of executable Eiffel \textsc{code} and of Eiffel \textsc{specification} (a total of $T$ specification lines, split into preconditions $P$, postconditions $Q$, frame specifications $F$, loop invariants $L$ and variants $V$, auxiliary annotations including ghost code $A$, and class invariants $C$); the \textsc{s}/\textsc{c} specification to code ratio (measured in tokens)\footnote{In accordance with common practices in verification competitions, we count \emph{tokens} for the \textsc{s}/\textsc{c} ratio; but we provide other measures in \emph{lines}, which are more naturally understandable.}; the lines of \textsc{Boogie} input (where \textit{tr} is the problem-specific translation code and \textit{bg} are the included background theory necessary for verification); the overall verification time (in bulk mode).
\AutoProof ran on a single core of a Windows 7 machine with a 3.5 GHz Intel i7-core CPU and 16 GB of memory, using Boogie v.~2.2.30705.1126 and Z3 v.~4.3.2 as backends.

\begin{table}
\begin{center}
\begin{scriptsize}
\setlength{\tabcolsep}{3pt}


\begin{tabular}{rl rrr r rrrrrrrr r@{.}l rr r}
\textsc{\#} & \textsc{name} & \textsc{\#C} & \multicolumn{2}{c}{\textsc{\#R}} & \textsc{code} & \multicolumn{8}{c}{\textsc{specification}} & \multicolumn{2}{c}{\textsc{s}/\textsc{c}} & \multicolumn{2}{c}{\textsc{Boogie}} & \multicolumn{1}{c}{\textsc{time} [s]} \\
&&&{\scriptsize \textit{co}}&{\scriptsize \textit{gh}}&& {\scriptsize $T$} & {\scriptsize $P$} & {\scriptsize $Q$} & {\scriptsize $F$} &{\scriptsize $L$} & {\scriptsize $V$} & {\scriptsize $A$} & {\scriptsize $C$}  &\multicolumn{2}{c}{}& {\scriptsize \textit{tr}} &{\scriptsize \textit{bg}} \\
\hline

\ref{p:arith} & \nref{p:arith} & 
1 & 6 & 0 & 
99 & 44 &
11 & 12 & 0 & 12 & 9 & 0 & 0 &
0&4 & 
927 & 579& 3.1 
\\

\ref{p:binsearch} & \nref{p:binsearch} &
1 & 4 & 1 & 
62 & 48 &
11 & 12 & 0 & 6 & 3 & 16 & 0 &
1&6 & 
965 & 1355 & 3.7
\\

\ref{p:sum-max} & \nref{p:sum-max} &
1 & 1 & 0 & 
23 & 12 &
3 & 2 & 1 & 4 & 0 & 2 & 0 &
1&0 & 
638 & 1355 & 3.9
\\

\ref{p:search-list} & \nref{p:search-list} &
2 & 5 & 1 & 
57 & 62 &
2 & 12 & 2 & 6 & 2 & 27 & 11 &
2&3 & 
931 & 1355 & 4.1
\\

\ref{p:twaymax} & \nref{p:twaymax} &
1 & 1 & 0 & 
23 & 12 &
2 & 4 & 0 & 4 & 2 & 0 & 0 &
2&3 & 
583 & 1355 & 3.0
\\

\ref{p:twaysort} & \nref{p:twaysort} &
1 & 2 & 0 & 
35 & 28 &
5 & 7 & 2 & 6 & 2 & 6 & 0 &
1&8 & 
683 & 1355 & 3.2
\\

\ref{p:dutchflag} & \nref{p:dutchflag} &
1 & 4 & 1 &
72 & 75 &
13 & 22 & 4 & 21 & 0 & 15 & 0 &
2&6 & 
1447 & 1355 & 4.1
\\

\ref{p:lcp} & \nref{p:lcp} &
2 & 2 & 0 &
40 & 28 &
4 & 7 & 0 & 6 & 2 & 9 & 0 &
1&0 & 
1359 & 1355 & 4.2
\\

\ref{p:rotation} & \nref{p:rotation} &
1 & 3 & 3 &
51 & 74 &
14 & 10 & 3 & 17 & 2 & 28 & 0 &
2&6 & 
1138 & 1355 & 4.1
\\

\ref{p:sorting} & \nref{p:sorting} &
1 & 9 & 6 &
177 & 219 &
31 & 38 & 9 & 56 & 5 & 80 & 0 &
2&6 & 
2302 & 1355 & 5.8 
\\

\cline{1-2}

\ref{p:iterator} & \nref{p:iterator} &
3 & 8 & 0 &
88 & 69 &
15 & 26 & 6 & 0 & 0 & 11 & 11 &
1&4 & 
1461 & 1355 & 8.9 
\\

\ref{p:subj-obs} & \nref{p:subj-obs} &
3 & 6 & 0 &
71 & 56 &
10 & 14 & 4 & 3 & 0 & 15 & 10 &
1&4 & 
1156 & 1355 & 4.4
\\

\ref{p:composite} & \nref{p:composite} &
2 & 5 & 3 &
54 & 125 &
19 & 18 & 5 & 0 & 2 & 72 & 9 &
4&3 & 
1327 & 1355 & 7.5 
\\

\ref{p:mclock} & \nref{p:mclock} &
3 & 7 & 0 &
63 & 61 &
9 & 14 & 5 & 0 & 0 & 26 & 7 &
1&8 & 
956 & 579 & 3.7 
\\

\ref{p:marriage} & \nref{p:marriage} &
2 & 5 & 0 &
45 & 50 &
12 & 11 & 3 & 0 & 0 & 19 & 5 &
2&3 & 
755 & 579 & 3.3
\\

\ref{p:dllist} & \nref{p:dllist} &
2 & 8 & 0 &
69 & 76 &
12 & 14 & 4 & 0 & 0 & 39 & 7 &
2&0 & 
891 & 579 & 4.4
\\

\ref{p:pip} & \nref{p:pip} &
2 & 5 & 1 &
54 & 111 &
23 & 18 & 6 & 0 & 1 & 56 & 7 &
3&9 & 
988 & 1355 & 5.8
\\

\cline{1-2}

\ref{p:closures} & \nref{p:closures} &
9 & 18 & 0 &
145 & 106 &
40 & 31 & 8 & 0 & 0 & 22 & 5 &
0&8 & 
2418 & 688 & 5.7
\\
 
\ref{p:strategy} & \nref{p:strategy} &
4 & 4 & 0 &
43 & 5 &
0 & 4 & 0 & 0 & 0 & 1 & 0 &
0&2 & 
868 & 579 & 3.3
\\

\ref{p:cmd} & \nref{p:cmd} &
6 & 8 & 0 &
77 & 32 &
4 & 14 & 2 & 0 & 0 & 10 & 5 &
0&7 & 
1334 & 579 & 3.3
\\

\ref{p:map} & \nref{p:map} &
1 & 8 & 0 & 
78 & 67 &
6 & 29 & 2 & 6 & 4 & 15 & 5 &
2&3 & 
1259 & 1355 & 4.1
\\

\ref{p:linked-queue} & \nref{p:linked-queue} &
4 & 13 & 1 &
121 & 101 &
11 & 26 & 1 & 0 & 0 & 48 & 15 &
1&5 & 
2360 & 1355 & 7.4
\\

\ref{p:tree-max} & \nref{p:tree-max} &
1 & 3 & 0 &
31 & 43 &
3 & 12 & 2 & 0 & 2 & 19 & 5 &
2&1 & 
460 & 1355 & 3.2
\\

\ref{p:ringbuffer} & \nref{p:ringbuffer} &
1 & 9 & 0 &
66 & 54 &
8 & 19 & 4 & 0 & 0 & 12 & 11 &
1&1 & 
1256 & 1355 & 4.4
\\

\ref{p:hashset} & \nref{p:hashset} &
5 & 14 & 5 &
146 & 341 &
45 & 39 & 10 & 20 & 2 & 197& 28 &
3&7 & 
3546 & 1355 & 13.7
\\

\ref{p:game1} & \nref{p:game1} &
4 & 8 & 0 &
165 & 93 &
16 & 13 & 4 & 31 & 3 & 10 & 16 &
1&2 & 
4044 & 1355 & 26.6
\\

\ref{p:game2} & \nref{p:game2} &
8 & 18 & 0 &
307 & 173 &
25 & 27 & 11 & 48 & 3 & 29 & 30 &
1&4 & 
7037 & 1355 & 54.2
\\

\hline
& \textit{total} &
72 & 184 & 22 &
2262 & 2165 &
354 & 455 & 98 & 246 & 44 & 784 & 184 &
1&9 & 
43089 & 1355 & 203.8

\end{tabular}
\end{scriptsize}
\end{center}
\caption{Verification of benchmark problems with \AutoProof.}
\label{tab:Solutions}
\end{table}

Given that we target full functional verification, our specification to code ratios are small to moderate, which demonstrates that \AutoProof's notation and methodology support concise and effective annotations for verification. 
Verification times also tend to be moderate, which demonstrates that \AutoProof's translation to Boogie is effective. 

To get an idea of the kinds of annotations required, \iflong and of their level of abstraction, \fi we computed the ratio $A/T$ of auxiliary to total annotations.
On average, 2.8 out of 10 lines of specification are auxiliary annotations;
the distribution is quite symmetric around its mean; auxiliary annotations are less than 58\% of the specification lines in all problems.
Auxiliary annotations tend to be lower level, since they outline intermediate proof goals which are somewhat specific to the way in which the proof is carried out.
Thus, the observed range of $A/T$ ratios seems to confirm how \AutoProof supports incrementality: complex proofs are possible but require more, lower level annotations.

\subsection{Open Challenges} \label{sec:open-challenges}

The collection of benchmark problems discussed in the previous sections shows, by and large, that \AutoProof is a state-of-the-art auto-active tool for the functional verification of object-oriented programs.
To our knowledge, no other auto-active verifier fully supports the complex reasoning about class invariants that is crucial to verify object-oriented pattern implementation such as \nref{p:subj-obs} and \nref{p:pip}.
It is important to remark that we're describing \emph{practical} capabilities of tools: other auto-active verifiers may support logics sufficiently rich to express the semantics of object-oriented benchmarks, but this is a far cry from automated verification that is approachable idiomatically at the level of a real object-oriented language.
Also, \AutoProof's performance is incomparable against that of interactive tools, which may still offer some automation but always have the option of falling back to asking users when verification gets stuck.

The flip side of \AutoProof's focus on supporting a real object-oriented language is that it may not be the most powerful tool to verify purely algorithmic problems.
The benchmarks have shown that \AutoProof still works quite well in that domain, and there are no intrinsic limitations that prevent from applying it to the most complex examples.
However, algorithmic verification is often best approached at a level that abstracts from implementation details (such as pointers and objects) and can freely use high-level constructs such as infinite maps and nondeterminism.
Verifiers such as Dafny~\cite{Dafny} and Why3~\cite{Why3}, whose input languages have been explicitly designed to match such abstraction level, are thus best suited for algorithmic verification, which is instead not the primary focus of \AutoProof.

Another aspect of the complexity vs.\ expressivity trade-off emerges when verifying realistic data structure implementations (or, more generally, object-oriented code as it is written in real-life projects). 
Tools such as Dafny offer a bare-bones framing methodology that is simple to learn (and to teach) and potentially very expressive; but it becomes unwieldy to reason about complicated implementations, which require to deal with an abundance of special cases by specifying each of them at a low level of detail---and annotational complexity easily leads to unfeasible verification.
\AutoProof's methodology is richer, which implies a steeper learning curve but also a variety of constructs and defaults that can significantly reduce the annotational overhead and whose custom Boogie translation offers competitive performance in many practical cases.

Given \AutoProof's goal of targeting a real programming language, there are few domain-specific features of the Eiffel language that are not fully supported but are used in practice in a variety of programs: reasoning in \AutoProof about strings and floating-point numbers
is limited by the imprecision of the verification models of such features.
For instance (see \autoref{sec:input-language}), floating point numbers are translated as infinite-precision reals; precise reasoning requires manually specifying properties of floating point operations.
Another domain deliberately excluded from \AutoProof so far is concurrent programming.
As a long term plan, we envision extending \AutoProof to cover these domains to the extent possible: precise functional verification of such features is still largely an open challenge for automated verification tools. 

A related goal of \AutoProof's research is verifying a fully-specified realistic data structure library---the first such verification carried out entirely with an auto-active tool.
This effort---one of the original driving forces behind designing \AutoProof's features---has been recently completed with the verification of the EiffelBase2 container library~\cite{EB2}.

\section{Discussion} \label{sec:conclusion}

How do \AutoProof's techniques and implementation generalize to other domains?
While Eiffel has its own peculiarities, it is clear that \AutoProof's techniques are applicable with little changes to other mainstream object-oriented languages such as Java and C\#; and that \AutoProof's architecture uses patterns that lead to proper designs in other object-oriented languages too.

A practically important issue is the input language, namely how to reconcile the conflicting requirements of supporting Eiffel as completely as possible and of having a convenient notation for expressing annotations necessary for auto-active verification. 
While Eiffel natively supports fundamental specification elements (pre- and postconditions and invariants), we had to introduce ad hoc notations, using naming conventions and dummy features, to express modifies clauses, ghost code, and other verification-specific directives in a way that is backward compatible with Eiffel syntax. 
We considered different implementation strategies, such as using a pre-processor or extending Eiffel's parser, but we concluded that being able to reuse standard Eiffel tools without modifying them is a better option in terms of reusability and compatibility (as the language and its tools may evolve), albeit it sacrifices a bit of notational simplicity.
This trade-off is reasonable whenever the goal is verifying programs in a real language used in practice; verifiers focused on algorithmic challenges would normally prefer ad hoc notations with an abstraction level germane to the tackled problems.

In future work, \AutoProof's architecture could integrate translations to back-end verifiers other than Boogie.
To this end, we could leverage verification systems such as Why3~\cite{Why3}, which generates verification conditions and discharges them using a variety of SMT solvers or other provers.

Supporting back-ends with different characteristics is one of the many aspects that affect the \emph{flexibility} of \AutoProof and similar tools.
Another crucial aspect is the quality of feedback in case of failed verification attempts, when users have to change the input to fix errors and inconsistencies, work around limitations of the back-end, or both.
As mentioned in \autoref{sec:usage}, \AutoProof incorporates heuristics that improve feedback. 
Another component of the EVE environment combines \AutoProof with automatic random testing and integrates the results of applying both~\cite{TFNM11-SEFM11}.
As future work we plan to further experiment with integrating the feedback of diverse code analysis tools (\AutoProof being one of them) to improve usability of verification.


\begin{thebibliography}{10}

\bibitem{SpecSharp}
M.~Barnett, M.~F{\"a}hndrich, K.~R.~M. Leino, P.~M{\"u}ller, W.~Schulte, and
  H.~Venter.
\newblock Specification and verification: the {Spec\#} experience.
\newblock {\em Commun. ACM}, 54(6):81--91, 2011.
\newblock \url{http://specsharp.codeplex.com/}.

\bibitem{Friends}
M.~Barnett and D.~A. Naumann.
\newblock Friends need a bit more: Maintaining invariants over shared state.
\newblock In {\em MPC}, pages 54--84, 2004.

\bibitem{KeY}
B.~Beckert, R.~H{\"a}hnle, and P.~H. Schmitt, editors.
\newblock {\em Verification of Object-Oriented Software: The {KeY} Approach},
  volume 4334 of {\em LNCS}.
\newblock Springer, 2007.

\bibitem{vc-cost11}
T.~Bormer et~al.
\newblock The {COST IC0701} verification competition 2011.
\newblock In {\em FoVeOOS}, volume 7421 of {\em LNCS}, pages 3--21. Springer,
  2012.
\newblock \url{http://foveoos2011.cost-ic0701.org/verification-competition}.

\bibitem{ESCJava2}
P.~Chalin, J.~R. Kiniry, G.~T. Leavens, and E.~Poll.
\newblock Beyond assertions: Advanced specification and verification with {JML}
  and {ESC}/{J}ava2.
\newblock In {\em FMCO}, LNCS, pages 342--363. Springer, 2006.
\newblock \url{http://kindsoftware.com/products/opensource/ESCJava2/}.

\bibitem{VCC}
E.~Cohen, M.~Dahlweid, M.~A. Hillebrand, D.~Leinenbach, M.~Moskal, T.~Santen,
  W.~Schulte, and S.~Tobies.
\newblock {VCC:} a practical system for verifying concurrent {C}.
\newblock In {\em TPHOLs}, volume 5674 of {\em LNCS}, pages 23--42. Springer,
  2009.
\newblock \url{http://vcc.codeplex.com/}.

\bibitem{OpenJMLPaper}
D.~Cok.
\newblock The {OpenJML} toolset.
\newblock In {\em NASA Formal Methods}, volume 6617. 2011.

\bibitem{Dij76}
E.~W. Dijkstra.
\newblock {\em A Discipline of Programming}.
\newblock Prentice Hall, 1976.

\bibitem{EB2}
{EiffelBase2}: A fully verified container library.
\newblock \url{https://github.com/nadia-polikarpova/eiffelbase2}, 2015.

\bibitem{Krakatoa}
J.-C. Filli{\^a}tre and C.~March{\'e}.
\newblock The {W}hy/{K}rakatoa/{C}aduceus platform for deductive program
  verification.
\newblock In {\em CAV}, volume 4590 of {\em LNCS}, pages 173--177. Springer,
  2007.
\newblock \url{http://krakatoa.lri.fr/}.

\bibitem{Why3}
J.-C. Filli{\^a}tre and A.~Paskevich.
\newblock {W}hy3 -- where programs meet provers.
\newblock In {\em ESOP}, volume 7792 of {\em LNCS}, pages 125--128. Springer,
  2013.
\newblock \url{http://why3.lri.fr/}.

\bibitem{vc-vstte12}
J.-C. Filli{\^a}tre, A.~Paskevich, and A.~Stump.
\newblock The 2nd verified software competition: Experience report.
\newblock In {\em COMPARE}, volume 873 of {\em CEUR Workshop Proceedings}.
  CEUR-WS.org, 2012.
\newblock \url{https://sites.google.com/site/vstte2012/compet}.

\bibitem{rotation}
C.~A. Furia.
\newblock Rotation of sequences: Algorithms and proofs.
\newblock \url{http://arxiv.org/abs/1406.5453}, June 2014.

\bibitem{Gof:1995}
E.~Gamma, R.~Helm, R.~Johnson, and J.~Vlissides.
\newblock {\em Design Patterns}.
\newblock Addison-Wesley, 1995.

\bibitem{vc-fm12}
M.~Huisman, V.~Klebanov, and R.~Monahan.
\newblock {V}erify{T}his verification competition.
\newblock \url{http://verifythis2012.cost-ic0701.org}, 2012.

\bibitem{JacobsSmansPiessens}
B.~Jacobs, J.~Smans, and F.~Piessens.
\newblock A quick tour of the {VeriFast} program verifier.
\newblock In {\em APLAS}, volume 6461 of {\em LNCS}, pages 304--311. Springer,
  2010.
\newblock \url{http://people.cs.kuleuven.be/~bart.jacobs/verifast/}.

\bibitem{KOA}
J.~R. Kiniry, A.~E. Morkan, D.~Cochran, F.~Fairmichael, P.~Chalin, M.~Oostdijk,
  and E.~Hubbers.
\newblock The {KOA} remote voting system: A summary of work to date.
\newblock In {\em TGC}, volume 4661 of {\em LNCS}, pages 244--262. Springer,
  2007.

\bibitem{vc-vstte10}
V.~Klebanov et~al.
\newblock The 1st verified software competition: Experience report.
\newblock In {\em FM}, volume 6664 of {\em LNCS}, pages 154--168. Springer,
  2011.
\newblock \url{https://sites.google.com/a/vscomp.org/main/}.

\bibitem{JML}
G.~T. Leavens, Y.~Cheon, C.~Clifton, C.~Ruby, and D.~R. Cok.
\newblock How the design of {JML} accommodates both runtime assertion checking
  and formal verification.
\newblock {\em Sci. Comput. Program.}, 55(1-3):185--208, 2005.

\bibitem{facj07}
G.~T. Leavens, K.~R.~M. Leino, and P.~M\"{u}ller.
\newblock Specification and verification challenges for sequential
  object-oriented programs.
\newblock {\em Formal Aspects of Computing}, 19(2):159--189, 2007.

\bibitem{BoogieManual}
K.~R.~M. Leino.
\newblock This is boogie 2.
\newblock Technical report, Microsoft Research, June 2008.
\newblock \url{http://research.microsoft.com/apps/pubs/default.aspx?id=147643}.

\bibitem{Dafny}
K.~R.~M. Leino.
\newblock {Dafny}: An automatic program verifier for functional correctness.
\newblock In {\em LPAR-16}, volume 6355 of {\em LNCS}, pages 348--370.
  Springer, 2010.
\newblock \url{http://research.microsoft.com/en-us/projects/dafny/}.

\bibitem{LeinoMoskal:UsableProgramVerification}
K.~R.~M. Leino and M.~Moskal.
\newblock Usable auto-active verification.
\newblock In {\em Usable Verification Workshop}.
  \url{http://fm.csl.sri.com/UV10/}, November 2010.

\bibitem{Dynamic}
K.~R.~M. Leino and P.~M{\"u}ller.
\newblock Object invariants in dynamic contexts.
\newblock In {\em ECOOP}, pages 491--516, 2004.

\bibitem{Clousot}
F.~Logozzo.
\newblock Our experience with the {CodeContracts} static checker.
\newblock In {\em VSTTE}, volume 7152 of {\em LNCS}, pages 241--242. Springer,
  2012.
\newblock \url{http://msdn.microsoft.com/en-us/devlabs/dd491992.aspx}.

\bibitem{OpenJML}
The {OpenJML} toolset.
\newblock \url{http://openjml.org/}, 2013.

\bibitem{PFM10-VSTTE10}
N.~Polikarpova, C.~A. Furia, and B.~Meyer.
\newblock Specifying reusable components.
\newblock In {\em VSTTE}, volume 6217 of {\em LNCS}, pages 127--141. Springer,
  2010.

\bibitem{semicola}
N.~Polikarpova, J.~Tschannen, C.~A. Furia, and B.~Meyer.
\newblock Flexible invariants through semantic collaboration.
\newblock In {\em FM}, volume 8442 of {\em LNCS}, pages 514--530. Springer,
  2014.

\bibitem{SAVCBS}
{SAVCBS} workshop series.
\newblock \url{http://www.eecs.ucf.edu/~leavens/SAVCBS/}, 2010.

\bibitem{Need}
A.~J. Summers, S.~Drossopoulou, and P.~M\"{u}ller.
\newblock The need for flexible object invariants.
\newblock In {\em IWACO}, pages 1--9. ACM, 2009.

\bibitem{Leon}
P.~Suter, A.~S. K\"{o}ksal, and V.~Kuncak.
\newblock Satisfiability modulo recursive programs.
\newblock In {\em SAS}, volume 6887 of {\em LNCS}, pages 298--315. Springer,
  2011.
\newblock \url{http://leon.epfl.ch/}.

\bibitem{sttt-paper}
J.~Tschannen, C.~A. Furia, and M.~Nordio.
\newblock {A}uto{P}roof meets some verification challenges.
\newblock {\em International Journal on Software Tools for Technology
  Transfer}, pages 1--11, February 2014.

\bibitem{TFNM11-SEFM11}
J.~Tschannen, C.~A. Furia, M.~Nordio, and B.~Meyer.
\newblock Usable verification of object-oriented programs by combining static
  and dynamic techniques.
\newblock In {\em SEFM}, volume 7041 of {\em LNCS}, pages 382--398. Springer,
  2011.

\bibitem{laser-paper}
J.~Tschannen, C.~A. Furia, M.~Nordio, and B.~Meyer.
\newblock Automatic verification of advanced object-oriented features: The
  {AutoProof} approach.
\newblock In {\em Tools for Practical Software Verification}, volume 7682 of
  {\em LNCS}, pages 133--155. Springer, 2012.

\bibitem{two-step-vstte}
J.~Tschannen, C.~A. Furia, M.~Nordio, and B.~Meyer.
\newblock Program checking with less hassle.
\newblock In {\em VSTTE 2013}, volume 8164 of {\em LNCS}, pages 149--169.
  Springer, 2014.

\bibitem{vsi-benchmarks}
B.~W. Weide, M.~Sitaraman, H.~K. Harton, B.~Adcock, P.~Bucci, D.~Bronish, W.~D.
  Heym, J.~Kirschenbaum, and D.~Frazier.
\newblock Incremental benchmarks for software verification tools and
  techniques.
\newblock In {\em VSTTE}, number 5295 in LNCS, pages 84--98. Springer, 2008.

\end{thebibliography}
\providecommand{\noopsort}[1]{}

\end{document}